\documentclass[11pt]{article}
\usepackage[utf8]{inputenc}
\usepackage{amsmath}
\usepackage{lipsum} % to generate some filler text
\usepackage{fullpage}
\usepackage{mathrsfs}
\usepackage{siunitx} 
\usepackage{xr}
\usepackage{amsmath}
\usepackage{booktabs}
\usepackage{array}
\usepackage{geometry}
\usepackage{longtable}
\usepackage{makecell}
\newcommand{\black}{\color{black}}
\usepackage{amsthm,cite,url,graphicx,booktabs,lipsum,color,bm,caption,subcaption,soul}
\usepackage{pifont,tikz,paralist,multirow,amssymb}
\usepackage{xifthen}
\usepackage{enumerate}
\usepackage{titlesec}
\usepackage{natbib} % load natbib package for bibliography
\usepackage{color}
\usepackage{tikz}
\usepackage{threeparttable}
\usepackage{hyperref} 

\usepackage{xcolor}
\definecolor{darkgreen}{RGB}{0,100,0}
\usepackage[normalem]{ulem}
\hypersetup{
    colorlinks=true,
    linkcolor=blue,
    urlcolor=blue,
    citecolor=blue
}
\usetikzlibrary{shapes.geometric,positioning,calc}
\usetikzlibrary{positioning}
\newtheorem{theorem}{Theorem}

\theoremstyle{definition}
 
\newtheorem{example}{Example}

\newtheorem{assumption}{Assumption}

\allowdisplaybreaks
\newcommand{\VC}{X}
\newcommand{\VCPimp}{X^{\dagger}}

\allowdisplaybreaks

\newcommand{\ormodel}[4]{
    \ifthenelse{\equal{#2}{C}}{
        \ifthenelse{\equal{#1}{0}}{
          Q_0(\VC, #3)
        }{
          Q_{#1}(\VC, #3; #4)
        }
    }{
        \ifthenelse{\equal{#2}{CP}}{
            \ifthenelse{\equal{#1}{0}}{
              Q_0(\VCPimp, #3)
            }{%
              Q_{#1}(\VCPimp, #3; #4)
            }%
            }{\textcolor{red}{\text{Invalid OR model specified}}}
    }
}

\newcommand{\psmodel}[4]{
    \ifthenelse{\equal{#2}{C}}{
        \ifthenelse{\equal{#1}{0}}{
          \pi_0(#3, \VC)
        }{
          \pi_{#1}(#3, \VC; #4)
        }
    }{
        \ifthenelse{\equal{#2}{CP}}{
            \ifthenelse{\equal{#1}{0}}{
              \pi_0(#3, \VC)
            }{%
              \pi_{#1}(#3, \VCPimp; #4)
            }%
            }{\textcolor{red}{\text{Invalid PS model specified}}}
    }
}

\def\pr{\textnormal{pr}}

\makeatletter
\newcommand*{\indep}{%
 \mathbin{%
  \mathpalette{\@indep}{}%
 }%
}
\newcommand*{\nindep}{%
 \mathbin{% % The final symbol is a binary math operator
  \mathpalette{\@indep}{\not}% \mathpalette helps for the adaptation
 }%
}
\newcommand*{\@indep}[2]{%
 \sbox0{$#1\perp\m@th$}% box 0 contains \indep symbol
 \sbox2{$#1=$}% box 2 for the height of =
 \sbox4{$#1\vcenter{}$}% box 4 for the height of the math axis
 \rlap{\copy0}% first \perp
 \dimen@=\dimexpr\ht2-\ht4-.2pt\relax
 \kern\dimen@
 {#2}%
 \kern\dimen@
 \copy0 % second \perp
}
\usepackage{enumitem}
\usepackage{xr}
\usepackage{dsfont}

\def\pr{\textnormal{pr}}

\usepackage{xr}
\makeatletter
\newcommand*{\addFileDependency}[1]{
  \typeout{(#1)}
  \@addtofilelist{#1}
  \IfFileExists{#1}{}{\typeout{No file #1.}}
}
\makeatother

 \newtheorem{innercustomgeneric}{\customgenericname}
\providecommand{\customgenericname}{}
\newcommand{\newcustomtheorem}[2]{%
  \newenvironment{#1}[1]
  {%
   \renewcommand\customgenericname{#2}%
   \renewcommand\theinnercustomgeneric{##1}%
   \innercustomgeneric
  }
  {\endinnercustomgeneric}
}
\newcustomtheorem{customthm}{Proposition}
\newcustomtheorem{customlemma}{Lemma}

\tikzset{
    varnode/.style={
        rectangle, 
        draw=black, 
        thick,
        fill=blue!10,
        minimum width=2.2cm,
        minimum height=0.8cm,
        text centered,
        font=\small
    },
    outcome/.style={
        rectangle,
        draw=black,
        thick,
        fill=orange!20,
        minimum width=2.5cm,
        minimum height=1cm,
        text centered,
        font=\small\bfseries
    },
    arrow/.style={
        ->,
        >=stealth,
        thick
    }
}
\tikzset{
    varnode/.style={
        ellipse,
        draw=black,
        thick,
        minimum width=2.5cm,
        minimum height=1cm,
        align=center,
        font=\small
    },
    arrow/.style={
        ->,
        >=stealth,
        thick
    }
}

\begin{document}
\hypersetup{linkcolor=black}
\title{\bf{\black Apportioning} Causal Responsibility of Two Risk Factors for an Adverse Outcome via Counterfactual Attribution}
  
 			\author{   Shanshan Luo\textsuperscript{1}, Yafang Deng\textsuperscript{1},  Qingyuan Zhao\textsuperscript{2},   and  Zhi Geng*\textsuperscript{1,3}  
       \\\\\textsuperscript{1} School of Mathematics and Statistics, \\Beijing Technology and Business University  \\[2mm] 
  	\textsuperscript{2}  {Statistical Laboratory,   University of Cambridge }   \\[2mm] 
  	\textsuperscript{3}  School of Mathematical Sciences, Peking University     	}

%  		\author{  Shanshan Luo\textsuperscript{1}, Yafang Deng\textsuperscript{1}, 	  and Zhi Geng\textsuperscript{1} \\\\ \vspace{0.15cm}
% \textsuperscript{1} School of Mathematics and Statistics,   Beijing Technology and Business University   }

\date{} % This removes the date
\maketitle  
\hypersetup{linkcolor=blue}
 
\begin{abstract}
Unlike traditional causal inference, which prospectively evaluates the effects of causes,
apportioning causal responsibility requires a retrospective assessment to deduce the
causes of an outcome that has already occurred. This paper proposes a quantitative
framework for apportioning causal responsibility between two binary risk factors that
jointly contribute to a realized adverse outcome. Ideally, knowing the individual's
latent causal type, defined by the potential outcomes under all possible exposure
combinations, would allow precise apportionment; however, these potential outcomes
cannot be simultaneously observed. We therefore define the average causal responsibility
of each risk factor as its expected responsibility over the distribution of latent
causal types. Under the assumptions of no confounding and monotonicity, we establish
nonparametric  identification of this metric when the type-specific responsibilities
satisfy a structural balance condition, and derive sharp bounds otherwise. We illustrate
the proposed framework using the classic example of lung cancer attributable to smoking
and asbestos exposures.
\end{abstract}
% \begin{abstract}
% Most existing literature on causal attribution focuses on prospectively evaluating the total effect of single or multiple causes on an outcome. Recent work has also examined retrospective posterior attribution given observed evidence,  including the outcome. However, there is limited discussion on the proportional contribution or responsibility share of each cause conditional on evidence, despite its critical importance in practice—for instance, in legal liability apportionment among multiple defendants, resource allocation in public health interventions, and compensation determination in tort cases. To address this gap, this paper proposes a framework for quantifying the responsibility or attribution proportion of each cause in multi-cause settings. Specifically, we develop a method to assess the attribution contribution, or responsibility, of each cause conditional on observed evidence, and quantify this contribution through posterior probability weighted averages. Under assumptions of sequential ignorability, monotonicity, and balance conditions on responsibility allocations, we establish point identification of attribution responsibility for each cause through observable equations, without requiring identification of all latent type probabilities. We illustrate the practical application of this framework through examples in occupational health and pediatric epidemiology, providing an effective tool for causal attribution analysis.
% \end{abstract}
  
\begin{keywords}
Attribution analysis; Causal inference; Causal responsibility; Causes of effects; Multiple interactive causes.
\end{keywords}

%  As usual, the \maketitle command creates the title and author/affiliations
%  display 

\maketitle
\section{Introduction}
\label{sec:intro}
 When multiple interacting factors contribute to an observed outcome, evaluating their individual contributions requires retrospective causal inference. This problem is pervasive across disciplines: it appears as disease attribution in epidemiology, credit assignment in economics, and liability apportionment in tort law. To illustrate this challenge, consider a classic scenario \citep{vanderweele2014tutorial}: when a worker with both a smoking history and asbestos exposure develops lung cancer, how should responsibility be apportioned between smoking and asbestos?

To answer this question, the evidence available in such cases is typically restricted to four conditional incidence rates. For instance, \citet{hilt1986previous} report a lung cancer risk of $\xi_{00} = 0.12\%$ under no exposure  (with subscripts indexing smoking and asbestos status, respectively), $\xi_{10} = 0.95\%$ under smoking alone, $\xi_{01} = 0.67\%$ under asbestos alone, and $\xi_{11} = 4.51\%$ under dual exposure. These four incidence rates suffice for the standard \emph{effects of causes} analysis in epidemiology: smoking raises the cancer risk by $\xi_{10} - \xi_{00} = 0.83\%$ and asbestos by $\xi_{01} - \xi_{00} = 0.55\%$. However, a retrospective question is fundamentally different: rather than asking how much smoking or asbestos elevates risk at the population level, it asks what share of responsibility each risk factor should bear, conditional on the disease having occurred. This retrospective question belongs to the class of \emph{causes of effects} \citep{dawid2014fitting,Dawid2022EffectsReview}, which requires reasoning about the counterfactual outcomes that this dually-exposed patient would have experienced under smoking alone, asbestos alone, or neither exposure.

The literature on the causes of effects problem has developed primarily in the single-cause setting \citep{pearl2000causality,dawid2000causal}. The probability of necessity (PN) \citep{pearl2000causality}, a statistical formalization of the but-for criterion in law \citep{wright1985causation}, provides  a criterion for whether a single exposure was necessary for the observed outcome. This single-cause framework, however, does not extend directly to the responsibility apportionment problem: both smoking and asbestos may individually satisfy the but-for criterion, so computing PN separately for each factor can yield necessity shares that sum to well over $100\%$, leaving the apportionment among them ambiguous. Recent work has extended causal attribution to multiple-cause settings \citep{lu2023evaluating, luo2025assessing}, but has focused on quantifying the posterior (i.e., conditional on the outcome having occurred) total effects of multiple causes  rather than apportioning responsibility among them.

In this paper, we develop a framework for apportioning causal responsibility between two risk factors for a realized binary outcome. Because the true causal mechanism that produced any specific individual's outcome is inherently unobservable, we define the average causal responsibility as a weighted average of type-specific responsibilities: the shares are determined by an apportionment rule that specifies, for each causal type, how responsibility should be split between two factors, while the weights correspond to the posterior probabilities of these types.  We show that the resulting average causal responsibility  is point-identified whenever the chosen rule satisfies a structural balance condition, and otherwise admits sharp nonparametric bounds.

For illustration, returning to the lung cancer example \citep{vanderweele2014tutorial}, suppose neither smoking nor asbestos can prevent lung cancer in any individual. Under this monotonicity assumption, Table~\ref{tab:no-mono-intro} lists the possible latent types of the dually-exposed worker and the corresponding responsibility shares. Since the worker's exact profile is unknown, a natural approach is to weight these type-specific responsibility shares by the probabilities of the profiles given that lung cancer occurred, yielding the overall responsibility borne by smoking and by asbestos.

\begin{table}[h]
    \centering
\caption{Responsibility shares borne by smoking and asbestos for each counterfactual profile, assigned $100\%$ to a sole active cause, $50\%$ to each of two active causes, or $0\%$ otherwise. The immune profile is excluded as incompatible with the observed disease.}
    \label{tab:no-mono-intro}
    \setlength{\tabcolsep}{8pt}
    \renewcommand{\arraystretch}{1.0}
    \resizebox{\textwidth}{!}{
    \begin{threeparttable}
    \begin{tabular}{c c @{\hspace{2em}} c c}
    \toprule
    Counterfactual profile & (Smoking, Asbestos) & Counterfactual profile & (Smoking, Asbestos) \\
    \midrule
    \makecell[c]{Cancer occurs iff both smoking \\ and asbestos exposure are present \\ (synergistic).} & $(50\%,\, 50\%)$ &
    \makecell[c]{Cancer occurs iff either smoking \\ or asbestos exposure is present \\ (parallel).} & $(50\%,\, 50\%)$ \\
    \addlinespace
    \makecell[c]{Cancer occurs iff smoking \\ is present (smoking-only).} & $(100\%,\, 0\%)$ &
    \makecell[c]{Cancer occurs iff asbestos \\ exposure is present (asbestos-only).} & $(0\%,\, 100\%)$ \\
    \addlinespace
    Cancer always occurs (doomed). & $(0\%,\, 0\%)$ &
    Cancer never occurs (immune). & Excluded \\
    \bottomrule
    \end{tabular}
    \end{threeparttable}}
\end{table}

With the incidence rates reported by \citet{hilt1986previous}, our main result in Section~\ref{sec:main-results} yields simple closed-form expressions for the average causal responsibilities of smoking and asbestos under the symmetric rule in Table~\ref{tab:no-mono-intro}:
\begin{align*}
\text{Smoking}: \frac{\xi_{11} + \xi_{10} - \xi_{01} - \xi_{00}}{2\xi_{11}} = 51.77\%, \quad
\text{Asbestos}: \frac{\xi_{11} + \xi_{01} - \xi_{10} - \xi_{00}}{2\xi_{11}} = 45.57\%.
\end{align*}
That is, smoking bears $51.77\%$ of the responsibility for the worker's lung cancer and asbestos bears $45.57\%$. The remaining $2.66\%$ cannot be explained by either exposure and is attributed to the worker's baseline vulnerability, such as genetic susceptibility or other unmeasured  factors.

% When a rule violates the balance condition, point identification no longer holds. The workplace-oriented rule in Table~\ref{tab:no-mono-intro}, which assigns a heavier share to occupational exposure, is one such example. Even so, our theoretical results yield sharp bounds of $[40.20\%, 41.42\%]$ for smoking and $[55.92\%, 57.14\%]$ for asbestos. The asbestos interval lies entirely above the $50\%$ threshold, so asbestos can be identified as the primary responsibility  despite only partial identification.
 \section{Notation and Definitions}
\label{sec:setup}  
\subsection{Setup and Motivating Examples}
\label{sec:setup-examples}
We consider two binary risk factors, $Z, M \in \{0, 1\}$, and a binary outcome, $Y \in \{0, 1\}$. Any other risk factors are subsumed into observed covariates $C$, which we condition on throughout and suppress for notational simplicity. We assume that $Z$ and $M$ do not causally affect one another.  As a running example, let $Z = 1$ denote smoking, $M = 1$ denote asbestos exposure, and $Y = 1$ denote the occurrence of lung cancer. Let $Y_{zm}$ denote the potential outcome under the joint intervention $(Z=z, M=m)$; for instance, $Y_{11}$ represents a worker's outcome under dual exposure, while $Y_{10}$ represents the outcome under smoking alone. We assume causal consistency, so that the observed outcome $Y$ equals the potential outcome corresponding to the realized exposures. Let $\mathcal{E}_{zm} = (Z=z, M=m, Y=1)$ denote the observed evidence of exposure and harm, and $\xi_{zm} = \pr(Y=1 \mid Z=z, M=m)$ the corresponding conditional incidence rate. Since responsibility apportionment is meaningful only when at least one factor is present, we focus on the three evidence configurations $\mathcal{E}_{11}$, $\mathcal{E}_{10}$, and $\mathcal{E}_{01}$.

Beyond the epidemiological example of smoking and asbestos, the following two examples, drawn from tort litigation and digital economics, illustrate different ways in which the problem of apportioning responsibility between two simultaneous causes arises.

\begin{example}[The  Doubling of Risk Criterion]
\label{ex:doubling-risk}
In \textit{Sienkiewicz v. Greif} \citep{sienkiewicz2011}, a claimant developed mesothelioma ($Y=1$) following exposure to two sources of asbestos: a general background exposure ($Z=1$) and an occupational exposure ($M=1$). The question is how to apportion liability for the disease between the two sources. A common rule for linking the disease to a single source requires the relative risk $\xi_{11}/\xi_{10}$ to be greater than two, known as the ``doubling of risk'' threshold. In this case, the relative risk was only $\xi_{11}/\xi_{10} \approx 1.18$, well below the threshold. Even so, the UK Supreme Court assigned full responsibility to the occupational exposure $M$, ruling that it had ``materially increased the risk'' of disease.
\end{example}

The challenge of causal attribution is not limited to assigning responsibility for adverse outcomes; the same logic applies to positive events, where the goal shifts from apportioning blame to allocating credit.

\begin{example}[The Last-Click Rule]
\label{ex:last-click-attribution}
In digital marketing, a user may convert ($Y=1$) following exposure to two sources of advertising: a display ad ($Z=1$) and a search ad ($M=1$). The question is how to allocate credit for the conversion between the two ads. A common practice for linking the conversion to a single source is ``last-click attribution'', which credits the last click for driving the outcome. In reality, empirical studies show that these channels interact synergistically \citep{xu2014path}, with display ads priming users to respond more strongly to subsequent search ads. Even so, standard metrics assign full credit to the last touchpoint, neglecting the contribution of the earlier exposure.
\end{example}

Although these examples differ in domain, they share a common structure: the observed outcome results from two simultaneous exposures, and the assignment of responsibility hinges on the latent causal mechanism linking those exposures to the outcome. We formalize this mechanism as a latent causal type in the next subsection.
\subsection{Causal Types and Probability of Types}
  
Let the random vector $T = (Y_{00}, Y_{01}, Y_{10}, Y_{11}) \in \{0,1\}^4$ denote an individual's causal type, with realization $t$. For notational convenience, we write $t$ in compact form, while occasionally using the expanded string notation $y_{00}y_{01}y_{10}y_{11}$ for interpretability. Each  realizations of $T$ corresponds to a distinct latent disease mechanism. For example, the type $T = 0001$ (with $Y_{11}=1$ only) represents a synergistic mechanism requiring both smoking and asbestos exposure, whereas $T = 0011$ (with $Y_{10}=Y_{11}=1$) indicates that smoking alone is sufficient, regardless of asbestos exposure. 

Under the consistency assumption, the observed evidence $\mathcal{E}_{zm} = (Z=z, M=m, Y=1)$ implies $Y_{zm}=1$. Therefore, only causal types satisfying $Y_{zm}=1$ are compatible with the observation. For instance, under the evidence $\mathcal{E}_{11}$, the feasible set reduces to the eight types with $Y_{11}=1$, i.e., $t \in \{0001, 0011, 0101, 0111, 1001, 1011, 1101, 1111\}$.  
Let $\pi(t) = \pr(T=t)$ denote the marginal probability of causal type $t$. We define the conditional \emph{probability of types} as
\begin{equation}
    \label{eq:posterior-type}
    \pi(t \mid \mathcal{E}_{zm}) = \pr(T=t \mid Z=z, M=m, Y=1),
\end{equation}
which represents the probability that an individual belongs to causal type $t$ given the observed evidence $\mathcal{E}_{zm}$. This formulation generalizes the notions of the probability of necessity \citep{pearl2000causality} and the probability of causation \citep{dawid2000causal} to a two-exposure setting.

\subsection{Type-Specific Responsibilities and Average Causal Responsibility}

While the probability of types $\pi(t \mid \mathcal{E}_{zm})$ in \eqref{eq:posterior-type} quantifies the likelihood of each causal type $T=t$ given the observed evidence $ \mathcal{E}_{zm}$, it does not specify  how causal responsibility should be {\black apportioned} between the two risk factors. To address this, we introduce the concept of \textit{type-specific responsibilities} to formalize the responsibility shares illustrated in Table~\ref{tab:no-mono-intro}.  For a given causal type $T = t$ under observed evidence $\mathcal{E}_{zm}$, let $R_{\scriptscriptstyle \mathrm{Z}}(t \mid \mathcal{E}_{zm})$ and $R_{\scriptscriptstyle \mathrm{M}}(t \mid \mathcal{E}_{zm})$ denote the causal responsibility apportioned to the realized exposures $Z=z$ and $M=m$, respectively, and let $R_0(t \mid \mathcal{E}_{zm})$ denote the residual responsibility unexplained by either cause, which is attributable to background factors. We require a complete partition of responsibility for each given  type $T=t$:   
\begin{equation*} 
    R_0(t \mid \mathcal{E}_{zm}) 
    + R_{\scriptscriptstyle \mathrm{Z}}(t \mid \mathcal{E}_{zm}) 
    + R_{\scriptscriptstyle \mathrm{M}}(t \mid \mathcal{E}_{zm}) = 1,  
\end{equation*} 
where each fraction satisfies $R_{\{\cdot\}} \in [0,1]$. {\black Naturally, we assume $R_{\scriptscriptstyle \mathrm{M}}(t \mid \mathcal{E}_{z0})=0$ for $z\in\{0,1\}$ and $R_{\scriptscriptstyle \mathrm{Z}}(t \mid \mathcal{E}_{0m})=0$  for $m\in\{0,1\}$, since a physically absent factor is causally inactive and therefore carries no responsibility.} These apportionments are typically pre-specified based on the substantive or legal context; specific strategies are detailed in Sections \ref{sec:main-results} and \ref{subsec:general-full-mono}.

Because a specific individual’s true causal type remains inherently unobservable, we define the \textit{average causal responsibility} $R_{j}(  \mathcal{E}_{zm})$ for each realized exposure $j\in\{\mathrm{Z},\mathrm{M}\}$ by taking the expectation of type-specific responsibilities over the probability of types: 
\begin{equation*} 
    R_{j}(  \mathcal{E}_{zm}) 
    = \sum_{t \in\{0,1\}^4} 
       R_{j}(t \mid   \mathcal{E}_{zm}) \, 
       \pi(t \mid   \mathcal{E}_{zm})  .
\end{equation*}
The residual background responsibility is then given by $R_0(  \mathcal{E}_{zm}) = 1 - R_{\scriptscriptstyle \mathrm{Z}}(  \mathcal{E}_{zm}) - R_{\scriptscriptstyle \mathrm{M}}(  \mathcal{E}_{zm})$, which structurally guarantees a complete partition of responsibility. 

To make these quantities concrete, consider again the lung cancer example under evidence $\black \mathcal{E}_{11}=(Z=1, M=1, Y=1)$. The terms $R_{\scriptscriptstyle \mathrm{Z}}(\black \mathcal{E}_{11})$ and $R_{\scriptscriptstyle \mathrm{M}}(\black \mathcal{E}_{11})$ represent the weighted average of responsibility apportioned to the smoking factor and the asbestos exposure, respectively, for this specific patient's lung cancer $(Y=1)$. The residual $R_0(\black \mathcal{E}_{11})$ captures the portion of responsibility unexplained by either smoking or asbestos, likely stemming from the patient's baseline vulnerability (e.g., genetics, age) and representing the proportion of responsibility the patient potentially bears themselves.

% \begin{remark}[From Factual Causation to Proportional Liability]
% \label{rmk:factual-causation}
% In tort law \citep{wright1985causation}, factual causation (cause-in-fact) traditionally seeks a deterministic link between risk factors and an outcome. In our proposed framework, the responsibility of each causal type $R_{j}(t \mid \black \mathcal{E}_{zm})$ in equation~\eqref{eq:efficiency} formalizes this concept by answering a specific counterfactual question: if the underlying disease mechanism $T=t$ were perfectly known, what share of responsibility should cause $j \in \{\mathrm{Z},\mathrm{M}\}$ bear in a physical or logical sense? By calculating the expectation in equation~\eqref{eq:identify-Qj}, our proposed metric $R_j(\black \mathcal{E}_{zm})$ connects the deterministic apportionment of factual causation $R_j(t \mid \black \mathcal{E}_{zm})$ with the {\black probabilistic uncertainty of the types $\pi(t \mid \black \mathcal{E}_{zm})$}. In this sense, our proposed metric provides a rigorous mathematical foundation for the doctrine of ``proportional liability'' in law.
% \end{remark}  
\section{Identification of {\black Average Causal Responsibility}}
\label{sec:main-results}

The average causal responsibility $R_j(\mathcal{E}_{zm})$ provides a quantitative basis for apportioning responsibility  through a retrospective view. Its identification, however, poses a statistical challenge: it depends on the distribution of latent causal types $\pi(t \mid \mathcal{E}_{zm})$, which is not identifiable from the observed conditional incidence rates $\xi_{zm}$ alone. This section studies the nonparametric identification of $R_j(\mathcal{E}_{zm})$, beginning with the standard assumption of no unmeasured confounding.

\begin{assumption}[No Confounding]
\label{assump:no-confounding}
$(Z,M) \indep (Y_{00}, Y_{01}, Y_{10}, Y_{11})$.
\end{assumption}

Assumption~\ref{assump:no-confounding} requires that the exposures are jointly independent of all potential outcomes, a condition that holds by design in randomized experiments. In observational studies \citep{rosenbaum1983assessing}, it can be relaxed to the standard conditional ignorability assumption given measured covariates $C$. As stated in Section~\ref{sec:setup-examples}, we suppress $C$ throughout the paper, with all probabilities understood to be conditional on $C$. Under Assumption~\ref{assump:no-confounding}, the observed conditional risk $\xi_{zm}$ nonparametrically identifies the   probability $\xi_{zm} = \pr(Y_{zm} = 1)$ for $(z, m) \in \{0, 1\}^2$.

We next consider the monotonicity assumption across the four potential outcomes, which rules out any protective effect from either risk factor.

\begin{assumption}[Monotonicity]
\label{assump:full-mono}
$Y_{00} \leq Y_{01} \leq Y_{11}$ and $Y_{00} \leq Y_{10} \leq Y_{11}$ almost surely.
\end{assumption}

In many legal and epidemiological contexts, risk factors are typically harmful, meaning their presence can only increase the risk of an adverse outcome. Monotonicity assumption \ref{assump:full-mono} therefore holds naturally in such settings \citep{vanderweele2014tutorial, rothman2008modern} and reduces the number of feasible causal types to the six listed in Table~\ref{tab:all_cause_combined}. Although this assumption narrows the range of feasible distributions, the probabilities of types $\pi(t \mid \mathcal{E}_{zm})$ are still not point-identified and remain governed by a single unidentified joint probability $\nu = \pr(Y_{01} = 1, Y_{10} = 1)$. As shown by \citet{darroch1994synergism},  under Assumption \ref{assump:full-mono}, the parameter $\nu$ must satisfy
\begin{equation}
\label{eq:nu-bounds}
\nu^{\ell} = \max\{\xi_{01} + \xi_{10} - \xi_{11},\, \xi_{00}\} \leq \nu \leq \nu^{u} = \min\{\xi_{01},\, \xi_{10}\}.
\end{equation}
Specifically, we have $\pi(1111 \mid \mathcal{E}_{zm}) =\xi_{00}/ \xi_{zm} $ for any $zm \in \{11, 10, 01\}$, and the remaining non-zero probabilities are
\begin{equation}
\label{eq:pi-bounds-all}
\begin{gathered}
    \pi(0111\mid\mathcal{E}_{zm})\in \big[ \nu^\ell - \xi_{00}, \, \nu^u - \xi_{00} \big] \Big/ \xi_{zm} ~~~\text{for }~~ zm \in \{11, 10, 01\}, \\[2pt]
    \pi(0101\mid\mathcal{E}_{z1})\in \big[ \xi_{01} - \nu^u, \, \xi_{01} - \nu^\ell \big] \Big/\xi_{z1} ~~~\text{for }~~ z \in \{0, 1\}, \\[2pt]
    \pi(0011\mid\mathcal{E}_{1m})\in \big[ \xi_{10} - \nu^u, \, \xi_{10} - \nu^\ell \big] \Big/ \xi_{1m} ~~~\text{for }~~ m \in \{0, 1\}, \\[2pt]
    \pi(0001\mid\mathcal{E}_{11})\in \big[ \xi_{11} - \xi_{10} - \xi_{01} + \nu^\ell, \, \xi_{11} - \xi_{10} - \xi_{01} + \nu^u \big] \Big/\xi_{11}.  
\end{gathered}
\end{equation}
with all other type probabilities equal to zero. Despite the lack of point identification of $\pi(t \mid \mathcal{E}_{zm})$, the average causal responsibilities may still be identifiable under suitable conditions on the type-specific responsibilities, as we now describe.

\begin{assumption}[Active-only Rule]
\label{symmetric-assump}
Given the observed evidence $\mathcal{E}_{zm}$, type-specific responsibilities are determined exclusively by which factors play an active causal role: a sole active factor assumes $100\%$ responsibility, two active factors share the responsibility equally at $50\%$ each, and any causally inactive factor bears $0\%$ responsibility, as specified in Table~\ref{tab:all_cause_combined}.
\end{assumption}

Assumption~\ref{symmetric-assump} embodies symmetry and fairness: responsibility is borne only by causally active factors, and active factors are treated symmetrically. To illustrate, consider the responsibility apportionment (smoking, asbestos) under the dual exposure evidence $\mathcal{E}_{11}$. For a patient of type $0001$ (synergistic), the cancer develops only under the joint presence of both exposures, so both factors are equally active and share responsibility equally ($50\%, 50\%$). For type $0011$, only smoking is causally active, since asbestos plays no role regardless of its presence, so full responsibility is assigned to smoking ($100\%, 0\%$); type $0101$ is symmetric, yielding $(0\%, 100\%)$. For type $0111$ (parallel), either factor alone is causally sufficient, so responsibility is again split equally ($50\%, 50\%$). For the doomed type $1111$, the cancer occurs regardless of either exposure, so neither factor is causally active ($0\%, 0\%$). The single-exposure cases are handled symmetrically: under $\mathcal{E}_{10}$, the active factor (smoking) takes $100\%$ responsibility for types whose causal mechanism is triggered by the sole realized exposure ($Z=1$), namely types $0011$ and $0111$, with $\mathcal{E}_{01}$ analogous.

\begin{table}[htbp]
    \centering
    \caption{The causal types and their responsibilities under single- and dual-exposure scenarios.}
    \label{tab:all_cause_combined}
    \resizebox{0.99995\textwidth}{!}{
    \begin{threeparttable}
    \centering
    \begin{tabular}{ccccc}
    \toprule
    \multirow{3}{*}{\makecell{Causal Types \\ $y_{00}y_{01}y_{10}y_{11}$}} &
    \multirow{3}{*}{\makecell[c]{Type Characteristic}} &
    \multicolumn{3}{c}{Type-Specific Responsibilities} \\
    \cmidrule(lr){3-5}
    & &
    \makecell[c]{$(R_{\scriptscriptstyle \mathrm{Z}}(\mathcal{E}_{11}),\, R_{\scriptscriptstyle \mathrm{M}}(\mathcal{E}_{11}))$} &
    \makecell[c]{$(R_{\scriptscriptstyle \mathrm{Z}}(\mathcal{E}_{10}),\, R_{\scriptscriptstyle \mathrm{M}}(\mathcal{E}_{10}))$} &
    \makecell[c]{$(R_{\scriptscriptstyle \mathrm{Z}}(\mathcal{E}_{01}),\, R_{\scriptscriptstyle \mathrm{M}}(\mathcal{E}_{01}))$} \\
    \midrule
    $0000$ &
    \makecell[c]{Cancer never occurs (immune).} &
    $\#$ & $\#$ & $\#$ \\
    \addlinespace[3pt]
    $0001$ &
    \makecell[c]{Cancer occurs iff both smoking and \\ asbestos exposure are present (synergistic).} &
    $(50\%,\, 50\%)$ & $\#$ & $\#$ \\
    \addlinespace[3pt]
    $0011$ &
    \makecell[c]{Cancer occurs iff smoking \\ is present (smoking-only).} &
    $(100\%,\, 0\%)$ & $(100\%,\, 0\%)$ & $\#$ \\
    \addlinespace[3pt]
    $0101$ &
    \makecell[c]{Cancer occurs iff asbestos \\ exposure is present (asbestos-only).} &
    $(0\%,\, 100\%)$ & $\#$ & $(0\%,\, 100\%)$ \\
    \addlinespace[3pt]
    $0111$ &
    \makecell[c]{Cancer occurs iff either smoking or \\ asbestos exposure is present (parallel).} &
    $(50\%,\, 50\%)$ & $(100\%,\, 0\%)$ & $(0\%,\, 100\%)$ \\
    \addlinespace[3pt]
    $1111$ &
    \makecell[c]{Cancer always occurs (doomed).} &
    $(0\%,\, 0\%)$ & $(0\%,\, 0\%)$ & $(0\%,\, 0\%)$ \\
    \bottomrule
    \end{tabular}
    \begin{tablenotes}
        \item \textit{Note}: $\#$ indicates types that are logically incompatible with the observed evidence.
    \end{tablenotes}
    \end{threeparttable}}
\end{table}

\begin{theorem}[Point Identification under Monotonicity]
\label{thm:full-mono}
Under Assumptions~\ref{assump:no-confounding}, \ref{assump:full-mono}, and~\ref{symmetric-assump}, the average causal responsibilities are point-identified as
\begin{equation}
\label{eq:full-mono-responsibilities}
\begin{gathered}
    R_0(\mathcal{E}_{11}) = \frac{\xi_{00}}{\xi_{11}}, \quad
    R_{\scriptscriptstyle \mathrm{Z}}(\mathcal{E}_{11}) = \frac{\xi_{11} + \xi_{10} - \xi_{01} - \xi_{00}}{2\xi_{11}}, \quad
    R_{\scriptscriptstyle \mathrm{M}}(\mathcal{E}_{11}) = \frac{\xi_{11} + \xi_{01} - \xi_{10} - \xi_{00}}{2\xi_{11}}, \\[0.5ex]
    R_0(\mathcal{E}_{10}) = \frac{\xi_{00}}{\xi_{10}}, \quad
    R_{\scriptscriptstyle \mathrm{Z}}(\mathcal{E}_{10}) = \frac{\xi_{10} - \xi_{00}}{\xi_{10}}, \quad
    R_{\scriptscriptstyle \mathrm{M}}(\mathcal{E}_{10}) = 0, \\[0.5ex]
    R_0(\mathcal{E}_{01}) = \frac{\xi_{00}}{\xi_{01}}, \quad
    R_{\scriptscriptstyle \mathrm{Z}}(\mathcal{E}_{01}) = 0, \quad
    R_{\scriptscriptstyle \mathrm{M}}(\mathcal{E}_{01}) = \frac{\xi_{01} - \xi_{00}}{\xi_{01}}.
\end{gathered}
\end{equation}
\end{theorem}

Theorem~\ref{thm:full-mono} shows that although the posterior probabilities $\pi(t \mid \mathcal{E}_{11})$ are only partially identified, the average causal responsibilities are nonparametrically point-identified, and the resulting expression under evidence $\mathcal{E}_{11}$ corresponds exactly to the result showcased in Section~\ref{sec:intro}. Identification obtains because the unidentifiable parameter $\nu$ in~\eqref{eq:nu-bounds} cancels out when type-specific responsibilities are aggregated under Assumption~\ref{symmetric-assump}, so the final apportionment relies solely on the observable conditional risks $\xi_{zm}$.

The formulas in \eqref{eq:full-mono-responsibilities} provide a straightforward rule for liability apportionment. The background responsibility $R_0(\mathcal{E}_{11}) = \xi_{00}/\xi_{11}$ isolates the baseline risk. The remaining responsibility is shared between the two active factors. Because responsibilities arising from interaction mechanisms (synergistic and parallel types) are divided equally, the difference in final responsibilities depends exclusively on their independent effects:
\begin{equation*}
    R_{\scriptscriptstyle \mathrm{Z}}(\mathcal{E}_{11}) - R_{\scriptscriptstyle \mathrm{M}}(\mathcal{E}_{11}) =  ({\xi_{10} - \xi_{01}})/{\xi_{11}}.
\end{equation*}
Thus, $R_{\scriptscriptstyle \mathrm{Z}}(\mathcal{E}_{11}) > R_{\scriptscriptstyle \mathrm{M}}(\mathcal{E}_{11})$ if and only if $\xi_{10} > \xi_{01}$. In our running example, smoking receives a greater share of responsibility than asbestos if and only if the absolute risk under smoking alone exceeds that under asbestos alone.
 
  Finally, in the single-exposure scenario, our framework naturally reduces to classical causal attribution results \citep{pearl2000causality,dawid2000causal}. Under the evidence $\mathcal{E}_{10}=({Z=1, M=0, Y=1})$, the responsibility of the active factor $Z=1$ is exactly $R{\scriptscriptstyle \mathrm{Z}}(\mathcal{E}_{10}) = (\xi_{10} - \xi_{00})/\xi_{10}$. This represents the proportion of liability apportioned solely to $Z$ when it is the only present exposure, which perfectly coincides with the probability of necessity   under standard monotonicity \citep{pearl2000causality,tian2000probabilities}, measuring the   probability that the outcome was actually caused by   $Z=1$.

\section{Extension to Arbitrary Type-Specific Responsibilities} 
\label{subsec:general-full-mono}
\subsection{Motivating Examples}
\label{subsec:discussion}
The type-specific responsibilities under Assumption~\ref{symmetric-assump} provide a natural baseline. However, they assign zero responsibility to the background factor for all causally active types, i.e., $R_0(t \mid \mathcal{E}_{11}) = 0$ for $t \in \{0001, 0011, 0101, 0111\}$. In practice, applications such as tort law or insurance claims may call for acknowledging that an individual's baseline vulnerability also contributes to the outcome. An apportionment rule may also reflect broader contextual considerations, such as assigning greater responsibility to background factors based on prior knowledge.
 \begin{table}[h]
    \centering
    \caption{Two alternative responsibility apportionments under single- and dual-exposure scenarios.}
    \label{tab:assignments-merged}
    \resizebox{\textwidth}{!}{
    \begin{threeparttable}
    \begin{subtable}{\textwidth}
    \centering
    \label{tab:assignment-balanced}
    \begin{tabular}{cccc}
    \toprule
    \multirow{2}{*}{\makecell{Causal Type \\ $y_{00}y_{01}y_{10}y_{11}$}} &
    \multicolumn{3}{c}{Balanced Type-Specific Responsibilities (Satisfying Subsequent Assumption \ref{assump:cancel})} \\
    \cmidrule(lr){2-4}
    &
    \makecell[c]{$(R_0(\mathcal{E}_{11}),\, R_{\scriptscriptstyle \mathrm{Z}}(\mathcal{E}_{11}),\, R_{\scriptscriptstyle \mathrm{M}}(\mathcal{E}_{11}))$} &
    \makecell[c]{$(R_0(\mathcal{E}_{10}),\, R_{\scriptscriptstyle \mathrm{Z}}(\mathcal{E}_{10}),\, R_{\scriptscriptstyle \mathrm{M}}(\mathcal{E}_{10}))$} &
    \makecell[c]{$(R_0(\mathcal{E}_{01}),\, R_{\scriptscriptstyle \mathrm{Z}}(\mathcal{E}_{01}),\, R_{\scriptscriptstyle \mathrm{M}}(\mathcal{E}_{01}))$} \\
    \midrule
    \makecell[c]{$0001$ \\ (synergistic)} &
    $(40\%,\, 30\%,\, 30\%)$ & $\#$ & $\#$ \\
    \addlinespace[3pt]
    \makecell[c]{$0011$ \\ (smoking-only)} &
    $(20\%,\, 60\%,\, 20\%)$ & $(20\%,\, 80\%,\, 0\%)$ & $\#$ \\
    \addlinespace[3pt]
    \makecell[c]{$0101$ \\ (asbestos-only)} &
    $(40\%,\, 10\%,\, 50\%)$ & $\#$ & $(20\%,\, 0\%,\, 80\%)$ \\
    \addlinespace[3pt]
    \makecell[c]{$0111$ \\ (parallel)} &
    $(20\%,\, 40\%,\, 40\%)$ & $(20\%,\, 80\%,\, 0\%)$ & $(20\%,\, 0\%,\, 80\%)$ \\
    \addlinespace[3pt]
    \makecell[c]{$1111$ \\ (doomed)} &
    $(80\%,\, 10\%,\, 10\%)$ & $(90\%,\, 10\%,\, 0\%)$ & $(90\%,\, 0\%,\, 10\%)$ \\ 
    \end{tabular}
    \end{subtable}

    \begin{subtable}{\textwidth}
    \centering
    \label{tab:assignment-workplace}
    \begin{tabular}{cccc}
    \toprule
    \multirow{2}{*}{\makecell{Causal Type \\ $y_{00}y_{01}y_{10}y_{11}$}} &
    \multicolumn{3}{c}{Workplace-oriented Type-Specific Responsibilities (Violating Subsequent Assumption \ref{assump:cancel})} \\
    \cmidrule(lr){2-4}
    &
    \makecell[c]{$(R_0(\mathcal{E}_{11}),\, R_{\scriptscriptstyle \mathrm{Z}}(\mathcal{E}_{11}),\, R_{\scriptscriptstyle \mathrm{M}}(\mathcal{E}_{11}))$} &
    \makecell[c]{$(R_0(\mathcal{E}_{10}),\, R_{\scriptscriptstyle \mathrm{Z}}(\mathcal{E}_{10}),\, R_{\scriptscriptstyle \mathrm{M}}(\mathcal{E}_{10}))$} &
    \makecell[c]{$(R_0(\mathcal{E}_{01}),\, R_{\scriptscriptstyle \mathrm{Z}}(\mathcal{E}_{01}),\, R_{\scriptscriptstyle \mathrm{M}}(\mathcal{E}_{01}))$} \\
    \midrule
    \makecell[c]{$0001$ \\ (synergistic)} &
    $(0\%,\, 40\%,\, 60\%)$ & $\#$ & $\#$ \\
    \addlinespace[3pt]
    \makecell[c]{$0011$ \\ (smoking-only)} &
    $(0\%,\, 80\%,\, 20\%)$ & $(60\%,\, 40\%,\, 0\%)$ & $\#$ \\
    \addlinespace[3pt]
    \makecell[c]{$0101$ \\ (asbestos-only)} &
    $(0\%,\, 0\%,\, 100\%)$ & $\#$ & $(40\%,\, 0\%,\, 60\%)$ \\
    \addlinespace[3pt]
    \makecell[c]{$0111$ \\ (parallel)} &
    $(0\%,\, 30\%,\, 70\%)$ & $(20\%,\, 80\%,\, 0\%)$ & $(20\%,\, 0\%,\, 80\%)$ \\
    \addlinespace[3pt]
    \makecell[c]{$1111$ \\ (doomed)} &
    $(90\%,\, 0\%,\, 10\%)$ & $(95\%,\, 5\%,\, 0\%)$ & $(85\%,\, 0\%,\, 15\%)$ \\
    \bottomrule
    \end{tabular}
    \end{subtable}
    \begin{tablenotes}
        \item \textit{Note}: $\#$ indicates types that are logically incompatible with the observed evidence.
    \end{tablenotes}
    \end{threeparttable}}
\end{table}

Table~\ref{tab:assignments-merged} presents two such alternatives, distinct from those in Table~\ref{tab:all_cause_combined}. Under dual exposure ($\mathcal{E}_{11}$), the balanced rule reserves a substantial share for background risks: the synergistic type ($0001$) assigns an equal $30\%$ share to each exposure and $40\%$ to the background, while the parallel type ($0111$) assigns more to each exposure ($40\%$ each), reflecting that independently sufficient factors bear greater responsibility than synergistic ones. For single-exposure scenarios ($\mathcal{E}_{10}$ and $\mathcal{E}_{01}$), both rules naturally assign $0\%$ responsibility to the absent factor, apportioning the remaining share strictly between the active exposure and a non-zero background. The workplace-oriented rule in Table~\ref{tab:assignments-merged}, by contrast, penalizes occupational asbestos exposure more heavily throughout, which is applicable when negligent asbestos exposure is judged more blameworthy than smoking. For example, under $\mathcal{E}_{11}$, even for the smoking-only type ($0011$), asbestos retains $20\%$ responsibility for creating a hazardous environment. Section~S7 of the Supplementary Material provides a symmetric smoking-oriented rule.
  
While these examples illustrate the flexibility of the framework, determining exact type-specific responsibilities in practice requires case-specific substantive analysis, and a full normative analysis of how such rules should be chosen is beyond the scope of this paper.

  \subsection{Point Identification under Balanced Responsibility} \label{partion-sec} Despite this flexibility, a natural statistical question is whether the average causal responsibility $R_j(\mathcal{E}_{zm})$ remains point-identified when the apportionment departs from the symmetric rule of Assumption~\ref{symmetric-assump}. The key insight is that identification only requires the type-specific responsibilities $R_j(t \mid \mathcal{E}_{zm})$ to satisfy a structural balance condition, which we formalize below.
%   \begin{assumption}[Balanced Responsibility]

% \label{assump:cancel}

% The type-specific responsibilities $R_j(y_{00}y_{01}y_{10}y_{11}  \mid \mathcal{E}_{zm})$ satisfy the following structural balance conditions:
% \begin{itemize}
%     \item[(i)] For the dual-exposure scenario ($\mathcal{E}_{11}$), the aggregate responsibility assigned to interactive types equals that of independent types for each factor $j \in \{\mathrm{Z}, \mathrm{M}\}$:
%     \begin{equation*}
%         R_j(0001 \mid \mathcal{E}_{11}) + R_j(0111 \mid \mathcal{E}_{11}) = R_j(0011 \mid \mathcal{E}_{11}) + R_j(0101 \mid \mathcal{E}_{11}).
%     \end{equation*}

%     \item[(ii)] For the single-exposure scenarios ($\mathcal{E}_{10}$ and $\mathcal{E}_{01}$), the responsibility of the active factor is invariant between its independent and parallel mechanisms:
%     \begin{equation*}
%         R_{\scriptscriptstyle \mathrm{Z}}(0011 \mid \mathcal{E}_{10}) = R_{\scriptscriptstyle \mathrm{Z}}(0111 \mid \mathcal{E}_{10}), \quad \text{and} \quad R_{\scriptscriptstyle \mathrm{M}}(0101 \mid \mathcal{E}_{01}) = R_{\scriptscriptstyle \mathrm{M}}(0111 \mid \mathcal{E}_{01}).
%     \end{equation*}

% \end{itemize}

% \end{assumption}
\begin{assumption}[Balanced Responsibility]
\label{assump:cancel}
The type-specific responsibilities $R_j(y_{00}y_{01}y_{10}y_{11}  \mid \mathcal{E}_{zm})$ satisfy the following structural balance conditions:
\begin{itemize}
    \item[(i)] For dual exposure  $\mathcal{E}_{11}$: $
        R_j(0001 \mid \mathcal{E}_{11}) + R_j(0111 \mid \mathcal{E}_{11}) = R_j(0011 \mid \mathcal{E}_{11}) + R_j(0101 \mid \mathcal{E}_{11})$ for $j \in \{\mathrm{Z}, \mathrm{M}\}$;  
  \item[(ii)] For single exposure $\mathcal{E}_{10}$: $
        R_{\scriptscriptstyle \mathrm{Z}}(0011 \mid \mathcal{E}_{10}) = R_{\scriptscriptstyle \mathrm{Z}}(0111 \mid \mathcal{E}_{10});$ 
  \item[(iii)] For single exposure $\mathcal{E}_{01}$: $
        R_{\scriptscriptstyle \mathrm{M}}(0101 \mid \mathcal{E}_{01}) = R_{\scriptscriptstyle \mathrm{M}}(0111 \mid \mathcal{E}_{01}).$
\end{itemize}
\end{assumption}

From a statistical perspective, Assumption~\ref{assump:cancel} imposes a structural constraint across the latent causal types. The left side sums the responsibility assigned to the \textit{interactive} types (the synergistic type $0001$ and the parallel type $0111$), while the right side sums that of the \textit{independent} types (the smoking-only type $0011$ and the asbestos-only type $0101$). This balance condition is strictly weaker than Assumption~\ref{symmetric-assump}. It allows asymmetric apportionments across types, requiring only aggregate balance between the interactive and independent groups. 
For instance, in the balanced asymmetric apportionment of
Table~\ref{tab:assignments-merged}, condition (i) holds for smoking
($Z$) despite the asymmetric apportionment across certain types:
$R_{\scriptscriptstyle \mathrm{Z}}(0001 \mid \mathcal{E}_{11}) +
R_{\scriptscriptstyle \mathrm{Z}}(0111 \mid \mathcal{E}_{11}) = 30\% +
40\% = 70\%$ matches $R_{\scriptscriptstyle \mathrm{Z}}(0011 \mid
\mathcal{E}_{11}) + R_{\scriptscriptstyle \mathrm{Z}}(0101 \mid
\mathcal{E}_{11}) = 60\% + 10\% = 70\%$, and likewise for asbestos
($M$), with $30\% + 40\% = 10\% + 60\% = 70\%$. Conditions (ii)-(iii) are similarly verified from Table~\ref{tab:assignments-merged}.

From a practical perspective, Assumption~\ref{assump:cancel} establishes an intuitive baseline for epidemiological attribution. Consider the lung cancer example. There are two broad ways the cancer can occur: smoking and asbestos may drive the disease together (interactive mechanisms), or one of them may trigger it on its own (independent mechanisms). The balance condition simply requires that, for each factor (smoking or asbestos), the total responsibility attributed to it in interactive scenarios equals its total responsibility in independent scenarios.
\begin{theorem}[Point Identification under Structural Balance]
\label{thm:full-mono-2}
Under Assumptions~\ref{assump:no-confounding}, \ref{assump:full-mono}, and~\ref{assump:cancel}, the average causal responsibilities $R_j(\mathcal{E}_{zm})$ are point-identified as follows. 

(i) For the dual-exposure scenario $\mathcal{E}_{11}$, $R_j(\mathcal{E}_{11})$ is given for $j \in \{0, \mathrm{Z}, \mathrm{M}\}$ by
\begin{equation*} 
\begin{aligned}
R_j(\mathcal{E}_{11}) &=  \frac{1}{\xi_{11}}\Big[ 
        \{R_j(1111\mid\mathcal{E}_{11}) - R_j(0111\mid\mathcal{E}_{11})\}\,\xi_{00} 
        + \{R_j(0101\mid\mathcal{E}_{11}) - R_j(0001\mid\mathcal{E}_{11})\}\,\xi_{01} \\
   &\qquad\quad + \{R_j(0011\mid\mathcal{E}_{11}) - R_j(0001\mid\mathcal{E}_{11})\}\,\xi_{10} 
        + R_j(0001\mid\mathcal{E}_{11})\,\xi_{11} 
\Big];
\end{aligned}
\end{equation*}

(ii) For the single-exposure scenario $\mathcal{E}_{10}$, $R_{\scriptscriptstyle \mathrm{M}}(\mathcal{E}_{10}) = 0$, and $R_j(\mathcal{E}_{10})$ is given for $j \in \{0, \mathrm{Z}\}$ by
\begin{equation*} 
\begin{aligned}
R_j(\mathcal{E}_{10}) &= \frac{1}{\xi_{10}} \Big[ \{R_j(1111 \mid \mathcal{E}_{10}) - R_j(0011 \mid \mathcal{E}_{10})\}\,\xi_{00} + R_j(0011 \mid \mathcal{E}_{10})\,\xi_{10} \Big];
\end{aligned}
\end{equation*}

(iii) For the single-exposure scenario $\mathcal{E}_{01}$, $R_{\scriptscriptstyle \mathrm{Z}}(\mathcal{E}_{01}) = 0$, and $R_j(\mathcal{E}_{01})$ is given for $j \in \{0, \mathrm{M}\}$ by
\begin{equation*} 
\begin{aligned}
R_j(\mathcal{E}_{01}) &= \frac{1}{\xi_{01}} \Big[ \{R_j(1111 \mid \mathcal{E}_{01}) - R_j(0101 \mid \mathcal{E}_{01})\}\,\xi_{00} + R_j(0101 \mid \mathcal{E}_{01})\,\xi_{01} \Big].
\end{aligned}
\end{equation*}
\end{theorem}

Theorem~\ref{thm:full-mono-2} shows that the balance condition in Assumption~\ref{assump:cancel} is sufficient for point identification of the average causal responsibilities. As a sanity check, substituting the active-only apportionment of Table~\ref{tab:all_cause_combined} into Theorem \ref{thm:full-mono-2}  recovers Theorem~\ref{thm:full-mono} as a special case. As a more substantive example, substituting the balanced apportionment from Table~\ref{tab:assignments-merged} under the dual-exposure scenario $\mathcal{E}_{11}$ yields
\begin{equation}
\label{eq:scneario-2}
\begin{gathered}
R_{0}(\mathcal{E}_{11}) = \frac{0.6\,\xi_{00} - 0.2\,\xi_{10} + 0.4\,\xi_{11}}{\xi_{11}}, \quad
R_{\scriptscriptstyle \mathrm{Z}}(\mathcal{E}_{11}) = \frac{-0.3\,\xi_{00} - 0.2\,\xi_{01} + 0.3\,\xi_{10} + 0.3\,\xi_{11}}{\xi_{11}}, \\[1.5pt]
R_{\scriptscriptstyle \mathrm{M}}(\mathcal{E}_{11}) = \frac{-0.3\,\xi_{00} + 0.2\,\xi_{01} - 0.1\,\xi_{10} + 0.3\,\xi_{11}}{\xi_{11}}.
\end{gathered}
\end{equation}
A notable feature of the formula \eqref{eq:scneario-2} is that, compared with the active-only formulas in~\eqref{eq:full-mono-responsibilities}, $R_0(\mathcal{E}_{11})$ now includes a positive contribution from $\xi_{11}$. Since $\xi_{11} \geq \xi_{10}$ and $\xi_{11} \geq \xi_{00}$ under monotonicity assumption  \ref{assump:full-mono}, this contribution is non-negligible, indicating that a larger share of responsibility is attributed to the patient's intrinsic vulnerability. This better aligns with apportionment in legal practice, where baseline vulnerability, such as age, chronic illness, or an unhealthy lifestyle, also serves as an important factor in assigning responsibility.
\subsection{Sharp Bounds with Arbitrary Type-Specific Responsibilities}

When the chosen apportionment rule is strongly asymmetric, the balance condition in Assumption~\ref{assump:cancel} may fail.  For example, under the workplace-oriented apportionment in Table~\ref{tab:assignments-merged}, smoking's responsibility for the interactive types sums to $40\% + 30\% = 70\%$, while its responsibility for the independent types sums to $80\% + 0\% = 80\%$; an analogous imbalance arises for asbestos. In such cases, the average causal responsibilities $R_j(\mathcal{E}_{zm})$ are no longer point-identified. They remain, however, sharply bounded by the observable risks.
\begin{theorem}[Sharp Bounds]
\label{thm:general-responsibility-bounds}
Under Assumptions~\ref{assump:no-confounding} and~\ref{assump:full-mono}, the average causal responsibility $R_j(\mathcal{E}_{zm})$ admits the sharp bounds 
\begin{equation*}
    R_j(\mathcal{E}_{zm}) \in \Big[ \alpha_{j}(\mathcal{E}_{zm}) + \min\!\left\{\beta_j(\mathcal{E}_{zm}) \nu^{\ell},  \beta_j(\mathcal{E}_{zm}) \nu^{u}\right\} ,\; \alpha_{j}(\mathcal{E}_{zm}) + \max\!\left\{\beta_j(\mathcal{E}_{zm}) \nu^{\ell},  \beta_j(\mathcal{E}_{zm}) \nu^{u}\right\} \Big],
\end{equation*} 
where $\nu^{\ell}$ and $\nu^{u}$ are the limits on the unidentified joint probability $\nu$ defined in~\eqref{eq:nu-bounds}, and the baseline $\alpha_j(\mathcal{E}_{zm})$ and imbalance coefficient $\beta_j(\mathcal{E}_{zm})$ are given by:

(i) For the dual-exposure scenario $\mathcal{E}_{11}$, $\alpha_j(\mathcal{E}_{11})$ and $\beta_j(\mathcal{E}_{11})$ are given for $j \in \{0, \mathrm{Z}, \mathrm{M}\}$ by 
\begin{equation*}
\begin{aligned}
    \alpha_{j}(\mathcal{E}_{11}) &= \frac{1}{\xi_{11}}\Big[ 
        \{R_j(1111\mid \mathcal{E}_{11}) - R_j(0111\mid \mathcal{E}_{11})\}\,\xi_{00} 
        + \{R_j(0101\mid \mathcal{E}_{11}) - R_j(0001\mid \mathcal{E}_{11})\}\,\xi_{01} \\
        &\qquad\quad + \{R_j(0011\mid \mathcal{E}_{11}) - R_j(0001\mid \mathcal{E}_{11})\}\,\xi_{10} 
        + R_j(0001\mid \mathcal{E}_{11})\,\xi_{11}  
    \Big], \\
    \beta_{j}(\mathcal{E}_{11}) &= \frac{1}{\xi_{11}}\Big[ R_j(0001 \mid \mathcal{E}_{11}) - R_j(0011 \mid \mathcal{E}_{11}) - R_j(0101 \mid \mathcal{E}_{11}) + R_j(0111 \mid \mathcal{E}_{11}) \Big];
\end{aligned}
\end{equation*}

(ii) For the single-exposure scenario $\mathcal{E}_{10}$, $\alpha_{\scriptscriptstyle \mathrm{M}}(\mathcal{E}_{10}) = \beta_{\scriptscriptstyle \mathrm{M}}(\mathcal{E}_{10}) = 0$, and for $j \in \{0, \mathrm{Z}\}$:
\begin{equation*}
\begin{aligned}
    \alpha_{j}(\mathcal{E}_{10}) &= \frac{1}{\xi_{10}} \Big[ \{R_j(1111 \mid \mathcal{E}_{10}) - R_j(0111 \mid \mathcal{E}_{10})\}\,\xi_{00} + R_j(0011 \mid \mathcal{E}_{10})\,\xi_{10} \Big], \\
    \beta_{j}(\mathcal{E}_{10}) &= \frac{1}{\xi_{10}} \Big[ R_j(0111 \mid \mathcal{E}_{10}) - R_j(0011 \mid \mathcal{E}_{10}) \Big];
\end{aligned}
\end{equation*}

(iii) For the single-exposure scenario $\mathcal{E}_{01}$, $\alpha_{\scriptscriptstyle \mathrm{Z}}(\mathcal{E}_{01}) = \beta_{\scriptscriptstyle \mathrm{Z}}(\mathcal{E}_{01}) = 0$, and for $j \in \{0, \mathrm{M}\}$:
\begin{equation*}
\begin{aligned}
    \alpha_{j}(\mathcal{E}_{01}) &= \frac{1}{\xi_{01}} \Big[ \{R_j(1111 \mid \mathcal{E}_{01}) - R_j(0111 \mid \mathcal{E}_{01})\}\,\xi_{00} + R_j(0101 \mid \mathcal{E}_{01})\,\xi_{01} \Big], \\
    \beta_{j}(\mathcal{E}_{01}) &= \frac{1}{\xi_{01}} \Big[ R_j(0111 \mid \mathcal{E}_{01}) - R_j(0101 \mid \mathcal{E}_{01}) \Big].
\end{aligned}
\end{equation*}
\end{theorem}

The coefficient $\beta_j(\mathcal{E}_{zm})$ in Theorem~\ref{thm:general-responsibility-bounds} controls the width of the bounds: it equals $|\beta_j(\mathcal{E}_{zm})|\,(\nu^{u} - \nu^{\ell})$, so apportionments with larger imbalance yield wider bounds. When $\beta_j(\mathcal{E}_{zm}) = 0$, Assumption~\ref{assump:cancel} (structural balance condition) is satisfied, and the bounds collapse to the point-identified value of Theorem~\ref{thm:full-mono-2}. Together, Theorems~\ref{thm:full-mono-2} and~\ref{thm:general-responsibility-bounds} cover any apportionment rule, with the precision of identification governed by $\beta_j(\mathcal{E}_{zm})$. To illustrate, consider the workplace-oriented apportionment in Table~\ref{tab:assignments-merged} under the dual-exposure scenario $\mathcal{E}_{11}$. The baseline parameters and imbalance coefficients are
\begin{equation*} 
\begin{gathered}
\alpha_{\scriptscriptstyle \mathrm{Z}}(\mathcal{E}_{11}) = \frac{-0.30\,\xi_{00} - 0.40\,\xi_{01} + 0.40\,\xi_{10} + 0.40\,\xi_{11}}{\xi_{11}}, \quad 
\beta_{\scriptscriptstyle \mathrm{Z}}(\mathcal{E}_{11}) = \frac{-0.10}{\xi_{11}}, \\[5pt]
\alpha_{\scriptscriptstyle \mathrm{M}}(\mathcal{E}_{11}) = \frac{-0.60\,\xi_{00} + 0.40\,\xi_{01} - 0.40\,\xi_{10} + 0.60\,\xi_{11}}{\xi_{11}}, \quad 
\beta_{\scriptscriptstyle \mathrm{M}}(\mathcal{E}_{11}) = \frac{0.10}{\xi_{11}}.
\end{gathered}
\end{equation*}
Substituting these values into Theorem~\ref{thm:general-responsibility-bounds} yields the sharp bounds on $R_{\scriptscriptstyle \mathrm{Z}}(\mathcal{E}_{11})$ and $R_{\scriptscriptstyle \mathrm{M}}(\mathcal{E}_{11})$.

Beyond their theoretical role, these sharp bounds are also informative in applied settings. Even when point identification fails, Theorem~\ref{thm:general-responsibility-bounds} delivers an explicit interval for the responsibility of each factor under the chosen apportionment rule, computed directly from the observable risks. Whenever this interval lies entirely on one side of a substantive threshold, a definitive conclusion about the relative importance of the factors can still be reached.
\section{Lung Cancer Attributable to Smoking and Asbestos: Numerical Examples}
\label{sec:numerical-examples}

In this section, we apply our framework to demonstrate its practical use in apportioning causal responsibility. To illustrate how our framework responds to different underlying disease mechanisms, we evaluate two scenarios: a real-world epidemiological dataset characterized by a substantial synergistic effect (Scenario I), and a synthetic dataset designed with a strong independent asbestos risk (Scenario II). Additional simulation studies are provided in Section~S8 of the Supplementary Material. %Let $Z$ and $M$ denote exposure to smoking and asbestos, respectively, with lung cancer $Y$ as the outcome ($1$ = present, $0$ = absent).

\subsection{Scenario I: Real-World Epidemiological Data}
\label{data:real}

Data from \citet{vanderweele2014tutorial} and \citet{hilt1986previous} yield the following conditional incidence rates:
\begin{equation*}
    \xi_{00} = 0.12\%, \quad \xi_{10} = 0.95\%, \quad \xi_{01} = 0.67\%, \quad \xi_{11} = 4.51\%.
\end{equation*}
Since $\xi_{00} \leq \min(\xi_{10}, \xi_{01}) $ and  $\max(\xi_{10}, \xi_{01}) \leq \xi_{11}$, the monotonicity condition in Assumption~\ref{assump:full-mono} is empirically supported.

Applying the inequalities \eqref{eq:pi-bounds-all}, we obtain the sharp bounds for the probabilities of types $\pi(t \mid \mathcal{E}_{zm})$ summarized in Table~\ref{tab:pi-bounds-scenario1}. The synergistic type ($0001$) dominates under dual-exposure $\mathcal{E}_{11}$, with $\pi(0001 \mid \mathcal{E}_{11}) \in [66.74\%, 78.94\%]$, indicating that lung cancer is primarily driven by the joint presence of both exposures. Under single exposure, the corresponding independent type dominates: the smoking-only type ($0011$) reaches up to $87.37\%$ under $\mathcal{E}_{10}$, and the asbestos-only type ($0101$) reaches up to $82.09\%$ under $\mathcal{E}_{01}$. This concentration in the synergistic type $\pi(0001 \mid \mathcal{E}_{11}) $ sets the basis for the responsibility allocations below.

\begin{table}[htbp]
\centering
\renewcommand{\arraystretch}{1.3}
\caption{Sharp bounds for the probabilities of     types $\pi(t \mid \mathcal{E}_{zm})$ under   three different evidence, for Scenario I (real-world data).}
\label{tab:pi-bounds-scenario1}
\resizebox{\textwidth}{!}{%
\begin{threeparttable}
\begin{tabular}{@{}lccccc@{}}
\toprule
Evidence & 
\makecell{Synergistic\\$\pi(0001\mid \mathcal{E}_{zm})$} & 
\makecell{Smoking-only\\$\pi(0011\mid \mathcal{E}_{zm})$} & 
\makecell{Asbestos-only\\$\pi(0101\mid \mathcal{E}_{zm})$} & 
\makecell{Parallel\\$\pi(0111\mid \mathcal{E}_{zm})$} & 
\makecell{Doomed\\$\pi(1111\mid \mathcal{E}_{zm})$} \\
\midrule
Dual Exposure ($\mathcal{E}_{11}$)   & $[66.74\%, 78.94\%]$ & $[6.21\%, 18.40\%]$ & $[0.00\%, 12.20\%]$ & $[0.00\%, 12.20\%]$ & $2.66\%$ \\
Smoking Only ($\mathcal{E}_{10}$)    & \# & $[29.47\%, 87.37\%]$ & \# & $[0.00\%, 57.89\%]$ & $12.63\%$ \\
Asbestos Only ($\mathcal{E}_{01}$)   & \# & \# & $[0.00\%, 82.09\%]$ & $[0.00\%, 82.09\%]$ & $17.91\%$ \\
\bottomrule
\end{tabular}
\begin{tablenotes}
    \item \textit{Note}: The symbol \# marks causal types logically incompatible with the observed evidence.
\end{tablenotes}
\end{threeparttable}%
}
\end{table}

We next evaluate the average causal responsibility $R_j(\mathcal{E}_{zm})$ under four apportionment rules: the active-only rule in Table~\ref{tab:all_cause_combined}, the balanced rule in Table~\ref{tab:assignments-merged}, the smoking-oriented rule in Section~S7 of the Supplementary Material, and the workplace-oriented rule in Table~\ref{tab:assignments-merged}. The estimation results are summarized in Table~\ref{tab:responsibility-scenario1}.

\begin{table}[h]
\centering
\renewcommand{\arraystretch}{1.3}
\caption{Average causal responsibilities $R_j(\mathcal{E}_{zm})$ across different evidence under four apportionment rules, for Scenario I (real-world data).}
\label{tab:responsibility-scenario1}
\resizebox{\textwidth}{!}{%
\begin{threeparttable}
\begin{tabular}{@{}llcccc@{}}
\toprule
& & \multicolumn{2}{c}{Point Identification ($\beta_j(\mathcal{E}_{zm}) = 0$)} & \multicolumn{2}{c}{Partial Identification ($\beta_j(\mathcal{E}_{zm}) \neq 0$)} \\
\cmidrule(lr){3-4} \cmidrule(l){5-6}
Evidence & Target Factor & Active-only & Balanced & Smoking-oriented & Workplace-oriented \\
\midrule
\multirow{3}{*}{$\mathcal{E}_{11}$}
 & Smoking ($R_Z$)    & $51.77\%$ & $32.55\%$ & $[61.15\%, 62.37\%]$ & $[40.20\%, 41.42\%]$ \\
 & Asbestos ($R_M$)   & $45.57\%$ & $30.07\%$ & $[35.23\%, 36.45\%]$ & $[56.19\%, 57.41\%]$ \\
 & Background ($R_0$) & $2.66\%$  & $37.38\%$ & $2.39\%$             & $2.39\%$ \\
\addlinespace
\multirow{3}{*}{$\mathcal{E}_{10}$}
 & Smoking ($R_Z$)    & $87.37\%$ & $71.16\%$ & $88.00\%$ & $[35.58\%, 58.74\%]$ \\
 & Asbestos ($R_M$)   & $0.00\%$  & $0.00\%$  & $0.00\%$  & $0.00\%$ \\
 & Background ($R_0$) & $12.63\%$ & $28.84\%$ & $12.00\%$ & $[41.26\%, 64.42\%]$ \\
\addlinespace
\multirow{3}{*}{$\mathcal{E}_{01}$}
 & Smoking ($R_Z$)    & $0.00\%$  & $0.00\%$  & $0.00\%$             & $0.00\%$ \\
 & Asbestos ($R_M$)   & $82.09\%$ & $67.46\%$ & $[51.94\%, 68.36\%]$ & $[51.94\%, 68.36\%]$ \\
 & Background ($R_0$) & $17.91\%$ & $32.54\%$ & $[31.64\%, 48.06\%]$ & $[31.64\%, 48.06\%]$ \\
\bottomrule
\end{tabular}
\begin{tablenotes}
    \item \textit{Note}: $\beta_j(\mathcal{E}_{zm}) = 0$ satisfies Assumption~\ref{assump:cancel}, yielding point identification (Theorem~\ref{thm:full-mono-2}); $\beta_j(\mathcal{E}_{zm}) \neq 0$ yields partial identification with sharp bounds (Theorem~\ref{thm:general-responsibility-bounds}).
\end{tablenotes}
\end{threeparttable}%
}
\end{table}

Under the active-only rule, applying~\eqref{eq:full-mono-responsibilities} to the dual-exposure stratum ($\mathcal{E}_{11}$) yields $51.77\%$ for smoking, $45.57\%$ for asbestos, and $2.66\%$ for background factors. The near-symmetric split reflects the dominance of the synergistic type, under which the active-only rule splits responsibility equally between the two exposures, so the point-identified result supports a roughly balanced apportionment of joint responsibility, with smoking bearing a slightly larger share. For single-exposure cases, the active-only rule naturally shifts most of the responsibility to the sole present exposure: applying~\eqref{eq:full-mono-responsibilities} yields $   87.37\%$ for smoking under $\mathcal{E}_{10}$, and $ 82.09\%$ for asbestos under $\mathcal{E}_{01}$.

The balanced rule illustrates how an epidemiological risk assessment might explicitly assign a larger share of responsibility to background factors. Under $\mathcal{E}_{11}$, the background share rises substantially to $37.38\%$, leaving smoking and asbestos with $32.55\%$ and $30.07\%$ respectively. Like the active-only case, this apportionment remains point-identified because the responsibility values satisfy the balance condition $\beta_{\scriptscriptstyle \mathrm{Z}}(\mathcal{E}_{zm}) = \beta_{\scriptscriptstyle \mathrm{M}}(\mathcal{E}_{zm}) = 0$ in Theorem~\ref{thm:full-mono-2}.

Under the smoking-oriented rule, point identification fails, but partial identification via Theorem~\ref{thm:general-responsibility-bounds} restricts smoking's responsibility under $\mathcal{E}_{11}$ to the sharp bounds $[61.15\%, 62.37\%]$. The bound width is remarkably narrow (about $1.2\%$), and the entire interval lies above the $50\%$ threshold commonly used to designate the primary contributing factor. Thus, investigators can conclude that smoking bears primary responsibility without requiring an exact estimate.

The workplace-oriented rule produces the opposite pattern by design. Point identification again fails, but the sharp bounds shift asbestos's responsibility to $[56.19\%, 57.41\%]$ and smoking's to $[40.20\%, 41.42\%]$. The asbestos interval now lies entirely above the $50\%$ threshold, identifying asbestos as the primary contributor under this rule.

\subsection{Scenario II: Synthetic Data with a Strong Asbestos Risk}

To explore a hypothetical epidemiological setting in which asbestos exposure serves as the dominant independent cause, we construct a synthetic dataset. Modifying only $\xi_{01}$ from Section~\ref{data:real} to strengthen the independent effect of asbestos, we consider the following conditional incidence rates:
\begin{equation*}
    \xi_{00} = 0.12\%, \quad \xi_{10} = 0.95\%, \quad \xi_{01} = 4.00\%, \quad \xi_{11} = 4.51\%.
\end{equation*}
This synthetic data is structured such that whenever $M = 1$, the disease risk is substantially elevated, while the monotonicity condition in Assumption~\ref{assump:full-mono} remains empirically supported.

The sharp bounds for the probabilities of types $\pi(t \mid \mathcal{E}_{zm})$ in Table~\ref{tab:pi-bounds-scenario2} reflect this structure. Under dual exposure, the asbestos-only type ($0101$) overwhelmingly dominates with $\pi(0101 \mid \mathcal{E}_{11}) \in [67.63\%, 78.94\%]$, suppressing the synergistic type $\pi(0111 \mid \mathcal{E}_{11}) $ to at most $11.31\%$. Unlike Scenario I, where the synergistic type was largest, here the asbestos-only type accounts for the largest proportion, which sets the basis for the responsibility allocations below.

\begin{table}[htbp]
\centering
\renewcommand{\arraystretch}{1.3}
\caption{Sharp bounds for the probabilities of     types $\pi(t \mid \mathcal{E}_{zm})$ under the three different evidence, for Scenario II (synthetic data).}
\label{tab:pi-bounds-scenario2}
\resizebox{\textwidth}{!}{%
\begin{threeparttable}
\begin{tabular}{@{}lccccc@{}}
\toprule
Evidence & 
\makecell{Synergistic\\$\pi(0001\mid \mathcal{E}_{zm})$} & 
\makecell{Smoking-only\\$\pi(0011\mid \mathcal{E}_{zm})$} & 
\makecell{Asbestos-only\\$\pi(0101\mid \mathcal{E}_{zm})$} & 
\makecell{Parallel\\$\pi(0111\mid \mathcal{E}_{zm})$} & 
\makecell{Doomed\\$\pi(1111\mid \mathcal{E}_{zm})$} \\
\midrule
Dual Exposure ($\mathcal{E}_{11}$)   & $[0.00\%, 11.31\%]$ & $[0.00\%, 11.31\%]$ & $[67.63\%, 78.94\%]$ & $[7.10\%, 18.40\%]$ & $2.66\%$ \\
Smoking Only ($\mathcal{E}_{10}$)    & \# & $[0.00\%, 53.68\%]$ & \# & $[33.68\%, 87.37\%]$ & $12.63\%$ \\
Asbestos Only ($\mathcal{E}_{01}$)   & \# & \# & $[76.25\%, 89.00\%]$ & $[8.00\%, 20.75\%]$ & $3.00\%$ \\
\bottomrule
\end{tabular}
\begin{tablenotes}
    \item \textit{Note}: The symbol \# marks causal types logically incompatible with the observed evidence.
\end{tablenotes}
\end{threeparttable}%
}
\end{table}
Following the setup in Section~\ref{data:real}, we evaluate the average causal responsibility $R_j(\mathcal{E}_{zm})$ under the same four apportionment rules. The results are reported in Table~\ref{tab:responsibility-scenario2}.

We begin with the point-identified cases. Satisfying the balance condition in Assumption \ref{assump:cancel}, both the active-only and balanced rules yield exact responsibilities using Theorem~\ref{thm:full-mono-2}. Under the active-only rule given $\mathcal{E}_{11}$, asbestos dominates at $82.48\%$ compared with $14.86\%$ for smoking, contrasting sharply with the synergistic balance observed in Scenario I. The balanced rule exhibits similar asymmetry: while background factors retain a $37.38\%$ share, the responsibility attributed to asbestos substantially exceeds that of smoking ($44.83\%$ relative to $17.78\%$).

In contrast, exact responsibilities are unidentified under the two oriented rules, but the framework still provides clear results using Theorem~\ref{thm:general-responsibility-bounds}. Although the smoking-oriented rule is designed to assign a heavier share of responsibility to smoking, its responsibility under $\mathcal{E}_{11}$ is bounded strictly below $50\%$ ($[32.33\%, 33.46\%]$). This firmly rules out smoking as the primary cause. Meanwhile, the workplace-oriented rule increases asbestos's share to $[86.43\%, 87.56\%]$. Therefore, under asbestos-dominant risks, the framework consistently identifies asbestos as the primary contributing factor across all rules and evidence configurations.
\begin{table}[t]
\centering
\renewcommand{\arraystretch}{1.3}
\caption{Average causal responsibilities $R_j(\mathcal{E}_{zm})$ across different evidence under four apportionment rules, for Scenario II (synthetic data).}
\label{tab:responsibility-scenario2}
\resizebox{\textwidth}{!}{%
\begin{threeparttable}
\begin{tabular}{@{}llcccc@{}}
\toprule
& & \multicolumn{2}{c}{Point Identification ($\beta_j(\mathcal{E}_{zm}) = 0$)} & \multicolumn{2}{c}{Partial Identification ($\beta_j(\mathcal{E}_{zm}) \neq 0$)} \\
\cmidrule(lr){3-4} \cmidrule(l){5-6}
Evidence & Target Factor & Active-only & Balanced & Smoking-oriented & Workplace-oriented \\
\midrule
\multirow{3}{*}{$\mathcal{E}_{11}$}
 & Smoking ($R_Z$)    & $14.86\%$ & $17.78\%$ & $[32.33\%, 33.46\%]$ & $[10.04\%, 11.18\%]$ \\
 & Asbestos ($R_M$)   & $82.48\%$ & $44.83\%$ & $[64.15\%, 65.28\%]$ & $[86.43\%, 87.56\%]$ \\
 & Background ($R_0$) & $2.66\%$  & $37.38\%$ & $2.39\%$             & $2.39\%$ \\
\addlinespace
\multirow{3}{*}{$\mathcal{E}_{10}$}
 & Smoking ($R_Z$)    & $87.37\%$ & $71.16\%$ & $88.00\%$ & $[49.05\%, 70.53\%]$ \\
 & Asbestos ($R_M$)   & $0.00\%$  & $0.00\%$  & $0.00\%$  & $0.00\%$ \\
 & Background ($R_0$) & $12.63\%$ & $28.84\%$ & $12.00\%$ & $[29.47\%, 50.95\%]$ \\
\addlinespace
\multirow{3}{*}{$\mathcal{E}_{01}$}
 & Smoking ($R_Z$)    & $0.00\%$  & $0.00\%$  & $0.00\%$             & $0.00\%$ \\
 & Asbestos ($R_M$)   & $97.00\%$ & $77.90\%$ & $[60.25\%, 62.80\%]$ & $[60.25\%, 62.80\%]$ \\
 & Background ($R_0$) & $3.00\%$  & $22.10\%$ & $[37.20\%, 39.75\%]$ & $[37.20\%, 39.75\%]$ \\
\bottomrule
\end{tabular}
\begin{tablenotes}
    \item \textit{Note}: $\beta_j(\mathcal{E}_{zm}) = 0$ satisfies Assumption~\ref{assump:cancel}, yielding point identification (Theorem~\ref{thm:full-mono-2}); $\beta_j(\mathcal{E}_{zm}) \neq 0$ yields partial identification with sharp bounds (Theorem~\ref{thm:general-responsibility-bounds}).
\end{tablenotes}
\end{threeparttable}%
}
\end{table}
% The two oriented apportionments, smoking-oriented and workplace-oriented, correspond to partial identification in Theorem~\ref{thm:general-responsibility-bounds} and demonstrate the robustness of the framework under alternative apportionment rules. In Scenario I ($\mathcal{E}_{11}$), the smoking-oriented rule elevates smoking's responsibility to $[61.15\%, 62.37\%]$, exceeding the $50\%$ threshold commonly used to designate the primary cause. The framework's strength is even more apparent in Scenario II: even under the smoking-oriented rule, which maximally favors smoking, smoking's responsibility under $\mathcal{E}_{11}$ is bounded within $[32.33\%, 33.46\%]$. Since the entire interval lies strictly below $50\%$, smoking can be firmly identified as not the primary cause, even under a rule deliberately calibrated to maximize smoking's share.  Consequently, under asbestos-dominant risks, the framework consistently identifies asbestos as the primary cause regardless of which apportionment rule is applied or whether dual or single exposure evidence is presented.
\section{Discussion}
In this paper, we proposed a formal framework for apportioning 
causal responsibility between two binary risk factors. Under 
monotonicity, average causal responsibilities are point-identified 
from observed conditional risks when type-specific responsibilities 
are balanced, and sharp, highly informative bounds are available 
for unbalanced apportionments. Applying the framework to lung 
cancer data on smoking and asbestos exposure, we showed that 
different apportionment rules yield substantively different 
allocations of responsibility, illustrating its practical value 
for legal and epidemiological applications.  
As discussed in 
Section~\ref{subsec:discussion}, the specification of type-specific 
responsibilities requires expert knowledge from the relevant domain 
(e.g., legal, digital economics, or epidemiological) and may rely 
on case-specific priors. Once this specification is made,  
the average causal responsibilities are fully determined from the 
observed data through the proposed framework, either by point 
identification or by sharp bounds.

For any specified type-specific responsibilities, an alternative 
perspective is that the identification of average causal 
responsibility ultimately depends on the joint distribution of 
potential outcomes $\pr(Y_{00}, Y_{01}, Y_{10}, Y_{11})$. Analogous 
to the extensive literature on identifying the joint distribution 
of potential outcomes in the single-cause setting, establishing 
broader identifiability conditions for this joint distribution in 
the two-cause setting is therefore a natural and important 
direction for future work. Promising avenues include partial 
identification approaches \citep{tian2000probabilities}, methods 
leveraging intermediate variables 
\citep{dawid2016bounding, rubinstein2025mediated}, proxy variable 
approaches \citep{kawakami2023identification,Shingaki2024}, and methods 
exploiting other auxiliary variables \citep{jiang2021identification,zhang2009likelihood}.

Our framework can also be extended in several structural directions. 
First, the methodology could be generalized to multiple causes. 
While we have established identifiability and sharp bounds for the 
two-cause setting, extending these results to three or more causes 
remains challenging, as the number of causal types and monotonicity 
configurations grows rapidly; establishing analogous identifiability 
conditions in this setting is a direction currently under 
investigation \citep{lu2023evaluating, luo2024assessing}. Second, 
the framework could be adapted to ordinal outcomes with more than 
two levels. Our current results focus on binary outcomes, and 
extending the type-based apportionment to ordinal outcomes requires 
redefining causal types and their type-specific responsibilities 
across multiple outcome levels \citep{lu2018treatment, 
zhang2025identifying}. Third, under weaker monotonicity assumptions, 
our identification results no longer apply directly; formulating 
appropriate type-specific responsibilities and developing partial 
identification strategies in this more general setting constitutes 
another valuable direction \citep{balke1994counterfactual, 
gabriel2024sharp, ben2024policy}. While these extensions lie beyond 
the scope of the present study, they represent natural progressions 
of the proposed methodology.
  \section*{Supplementary Material}

The Supplementary Material is organized as follows. Section~S7 introduces a supplementary smoking-oriented apportionment rule. Section~S8 reports simulation studies evaluating the finite-sample performance of the proposed estimators. Section~S9 provides the technical proofs.

\section*{Acknowledgments}
We are grateful to Philip Dawid for his helpful discussions on the formal definition of responsibility based on counterfactuals,  to Simon Deakin for providing  valuable legal examples, and to Martina Scauda for her helpful conversations on problems related to legal responsibility. This work was supported by the National Natural Science Foundation of China (12401378), the Young Elite Scientists Sponsorship Program of the Beijing High Innovation Plan (NO.20250864), the Beijing Key Laboratory of Applied Statistics and Digital Regulation, the BTBU Digital Business Platform Project by BMEC, and the BTBU Research Foundation for Youth Scholars (BRFYS2025). Dr. Shanshan Luo would like to thank the Isaac Newton Institute for Mathematical Sciences, Cambridge, for support and hospitality during the programme Causal inference: From theory to practice and back again, where work on this paper was undertaken. This work was supported by EPSRC grant EP/Z000580/1.

 \bibliographystyle{apalike}
					\bibliography{mybib} 

\end{document}